%% file: sofi2026algqst_eusipco.tex
\documentclass[conference]{IEEEtran}
\IEEEoverridecommandlockouts
\usepackage{graphicx}
\usepackage{dcolumn}
\usepackage{textcomp}
\usepackage{xcolor}
\usepackage{amsmath,amssymb,amsfonts}
\usepackage{subfigure}
\usepackage{hyperref} 
\usepackage{pgfplots}
\usepackage{wrapfig}
\usetikzlibrary{pgfplots.groupplots}
\usepackage[style=ieee,backend=biber,doi=false,isbn=false,url=false]{biblatex}
\usepackage{siunitx}
\usepackage{comment}
\pgfplotsset{compat=1.17}
\usepackage{lipsum}
\def\BibTeX{{\rm B\kern-.05em{\sc i\kern-.025em b}\kern-.08em
    T\kern-.1667em\lower.7ex\hbox{E}\kern-.125emX}}

\usepackage{algcompatible}
\usepackage[linesnumbered, lined, boxed]{algorithm2e}
\RestyleAlgo{ruled}
\SetKwComment{tcc}{/*}{*/}
\SetNoFillComment
\DontPrintSemicolon
\SetAlgoCaptionLayout{flushleft}
\SetAlgoVlined  
\SetCommentSty{mycommfont}

\usepackage{extramath}

\newcommand{\col}[1]{\text{col}\left( #1 \right)}  
\newcommand{\iset}[1]{\mathbf{\mathtt{#1}}} 

\newcommand{\midtiny}{\fontsize{7.3pt}{8pt}\selectfont}

\DeclareMathOperator*{\argmin}{arg\,min} 
\DeclareMathOperator*{\argmax}{arg\,max} 

\let\oldrho\rho
\renewcommand{\rho}{\bm{\oldrho}}

\usepackage{etoolbox}
\AtBeginBibliography{\midtiny} 

\begin{document}

\title{Tomography by Design: An Algebraic Approach to Low-Rank Quantum States \thanks{\scriptsize This work was supported by the Flemish Government's AI Research Program and KU Leuven Internal Funds (iBOF/23/064, C14/22/096). Shakir Showkat Sofi, Charlotte Vermeylen, and Lieven De Lathauwer are affiliated with Leuven.AI - KU Leuven institute for AI, B-3000, Leuven, Belgium.}
}

\author{\IEEEauthorblockN{Shakir Showkat Sofi}%
\IEEEauthorblockA{\textit{Dept. Electrical Engineering (ESAT)} \\%
\textit{KU Leuven}, Kortrijk/Leuven, Belgium \\%
shakirshowkat.sofi@kuleuven.be}%
\and
\IEEEauthorblockN{Charlotte Vermeylen}%
\IEEEauthorblockA{\textit{Dept. Electrical Engineering (ESAT)} \\%
\textit{KU Leuven}, Leuven, Belgium \\%
charlotte.vermeylen@kuleuven.be%
}
\and
\IEEEauthorblockN{Lieven De Lathauwer}%
\IEEEauthorblockA{\textit{Dept. Electrical Engineering (ESAT)} \\%
\textit{KU Leuven}, Kortrijk/Leuven, Belgium  \\ %
lieven.delathauwer@kuleuven.be}%
}
\maketitle 
\begin{abstract}
We present an algebraic algorithm for quantum state tomography that leverages measurements of certain observables to estimate structured entries of the underlying density matrix. Under low-rank assumptions, the remaining entries can be obtained solely using standard numerical linear algebra operations.  The proposed algebraic matrix completion framework applies to a broad class of generic, low-rank mixed quantum states and, compared with state-of-the-art methods, is computationally efficient while providing deterministic recovery guarantees.
\end{abstract}


\section{\label{sec:intro} Introduction}
The state of a quantum system is fully described by its density matrix---a Hermitian, positive-semidefinite (PSD) operator with unit trace---that provides a unified representation of both pure and mixed quantum states (probabilistic mixtures of pure states). Its diagonal entries represent state populations (probabilities), while its off-diagonal entries encode quantum coherences between states \cite{nielsen2010quantum}. Accurately determining this matrix is a key challenge in quantum science, as many tasks, such as benchmarking and verification of quantum hardware, and fidelity estimation, essentially involve its estimation. This matrix is estimated by a technique known as quantum state tomography (QST) \cite{nielsen2010quantum, cramer2010efficient, gross2010csqst, liu2012csqst}. QST works by performing an informationally complete set of measurements on a large ensemble of identically prepared quantum systems. The collected data is subsequently used to reconstruct the density matrix. However, a major obstacle is the exponential growth in parameters (and correspondingly in measurements and reconstruction complexity) as the system grows, an obstacle known as the ``curse of dimensionality." \par

Physically relevant states occupy only a tiny, structured subset of Hilbert space, constrained by locality and finite-complexity dynamics that limit correlations and entanglement \cite{eisert2010colloquium, gross2010csqst, liu2012csqst}. This subset includes pure or low-entropy states, low-energy states of local Hamiltonians, product states, and structured states in which only a subset of entries carry valuable information.  In such cases, the density matrix has low rank. Exploiting these low-rank or structural constraints can tame the curse of dimensionality. \par

\paragraph*{Related Work}{When the density matrix is known to be low rank or close to a low-rank matrix, {\em low-rank} QST exploits the structure to reconstruct the state from fewer measurements. Rank information can be incorporated either implicitly, by promoting low rank through a rank-minimization objective, implemented via the nuclear norm as a convex surrogate, under measurement constraints \cite{gross2010csqst, liu2011universal, liu2012csqst}, or explicitly, by optimizing a fixed-rank factorized model, as in Burer–Monteiro–type formulations \cite{burer2003lrsdp, kyrillidis2018provable}. If the density matrix is instead viewed as a high-order tensor of low rank in the tensor sense, efficient and scalable tensor network representations—such as matrix product states (tensor trains)—can be used to estimate it \cite{cramer2010efficient, verstraete2007MPS, oseledets2010tensortrain, lanyon2017efficient, sofi2025bttqst}. \par
Another line of research avoids full QST and focuses on learning only specific aspects of the underlying density matrix. For example, {\em permutationally invariant} QST efficiently reconstructs the permutationally invariant part of the underlying quantum state, with the number of measurement settings scaling quadratically, and is well suited for states that are close to being permutationally symmetric, such as Dicke or spin-squeezed states \cite{toth2010permutationally, moroder2012permutationally}. As another example, {\em shadow tomography} constructs a compact classical representation known as ``classical shadow"---a collection of randomized sketches of the density matrix---obtained via random unitary transformations, allowing many observables (or arbitrary linear functions) to be estimated from the compact representation with measurement complexity that scales logarithmically in the number of observables \cite{aaronson2018shadow, huang2020predicting}. \par
Extending the idea of learning partial information, {\em selective} QST focuses on directly estimating only a chosen subset of density-matrix entries, enabling targeted estimation without reconstructing the full state \cite{baldwin2016strictly, calderaro2018direct, feng2021direct, morris2019selective}. This approach can exploit additional structure: for instance, if the selected entries follow a specific low-rank--exploitable pattern, they can uniquely determine the global low-rank density matrix. Furthermore, these entries can be obtained via shadow tomography, allowing scalable estimation even of the full state \cite{wang2025direct}. 

}

\paragraph*{Contributions}{In this work, we further develop selective QST by introducing a novel algebraic QST method, based solely on standard numerical linear algebra (NLA) operations, for efficient recovery of the full density matrix from a subset of structured entries. We characterize such entries to establish unique and deterministic recovery guarantees, and provide subspace error bounds. We compare our method with state-of-the-art (SOTA) techniques and show that it is computationally efficient while achieving competitive accuracy in reconstructing mixed low-rank quantum states. Finally, we touch on how scalability can be improved by estimating structured entries via shadow tomography, combining low‑rank  and shadow tomography.
}
\subsection{\label{sec:notation} Preliminaries and Notation}
Scalars, vectors, and matrices are denoted by lowercase letters, bold lowercase letters, and bold uppercase letters, respectively; that is, $x$, $\vec{x}$, and $\mat{X}$. For $\mat{X} \in \mathbb{C}^{D \times D}$, the $(r,c)$-th entry is $x_{r c} = \mat{X}(r, c),$  and the $k$-th column is denoted by $\vec{x}_{:k} = \mat{X}(:,k)$. The trace, Frobenius norm, and nuclear norm are $\operatorname{Tr}(\mathbf{X})$, $\|\mathbf{X}\|_{F}$, and $\|\mathbf{X}\|_\star$.
Cartesian, Hadamard and Kronecker products are denoted by $\times$, $\ast$ and $\otimes$, respectively. Subspaces are denoted by uppercase letters; the dimension of a subspace $X$ is written $\dim(X)$. The (chordal) distance between $X$ and its estimate $X^\star$ is defined as 
\(
d_c(X, X^\star)
=
\frac{1}{\sqrt{2}}
\left\| \mat{P}_{X}-\mat{P}_{X^\star} \right\|_F ,
\)
where $\mathbf{P}_{X}$ and $\mathbf{P}_{X^\star}$ denote the orthogonal projection matrices onto $X$ and $X^\star$, respectively. A density matrix is denoted by \(\rho\). The space of density matrices is denoted by $\mathcal{S}$, where $\mathcal{S} = \{ \rho \in \mathbb{C}^{D \times D} \mid \rho = \rho^{\mathrm{H}}, ~ \rho \succeq 0,~ \operatorname{Tr}(\rho) = 1\}$. Sets are denoted by bold lowercase typewriter letters. The cardinality of a set $\iset{r}$ is denoted by $|\iset{r}|.$ 
The submatrix $\mat{X}_{\mathrm{obs}}^{(l)}=\mat{S}_{\iset{r}_{l}}^{\top}\mat{X}\mat{S}_{\iset{c}_{l}}=\mat{X}(\iset{r}_{l},\iset{c}_{l})\in\mathbb{C}^{|\iset{r}_{l}|\times |\iset{c}_{l}|}$ contains entries of $\mat{X}$ at the indices $(r,c)\in \iset{r}_{l}\times \iset{c}_{l}$, where $\iset{r}_{l}, \iset{c}_{l} \subseteq[D]=\{1,\ldots, D\}$, and $\mat{S}_{\iset{r}_{l}}^{\top} \in \{0, 1\}^{|\iset{r}_{l}| \times D}$ and $\mat{S}_{\iset{c}_{l}} \in \{0, 1\}^{D \times |\iset{c}_{l}|}$ are the respective row and column selection matrices. If $\iset{r}_{l}=\{r\}$ and $\iset{c}_{l}=\{c\}$, the observed entry is $x_{r c} = \vec{s}_{r}^{\top} \mat{X}\vec{s}_{c}$. A submatrix $\mat{X}^{(l)}_{\mathrm{obs}}$ is called \emph{isorank} if it has the same rank as the full matrix, i.e., 
$\operatorname{rank}(\mat{X}^{(l)}_{\mathrm{obs}})=\operatorname{rank}(\mat{X})$.\par

A \emph{qubit} is the basic unit of quantum information, capable of being in a superposition of $0$ and $1$. The \emph{Pauli spin matrices},  $\{\mathbf{I}_2, \bm{\sigma}_x,  \bm{\sigma}_y,  \bm{\sigma}_z\},$ provide a standard basis for describing and measuring qubit states.

\subsection{Outline}
Section~\ref{sec:learning} reviews learning Hermitian PSD matrices from partial data. Section~\ref{sec:algqst} outlines our estimation approach. Sections~\ref{sec:alg}–\ref{sec:uniq} give the algorithm and uniqueness results. Section~\ref{sec:exp} presents experiments, and Section~\ref{sec:con} concludes.

\section{Low-rank QST formulation}
\label{sec:learning}
QST can be formulated as an inverse problem. For an $N$-qubit system with density matrix $\rho \in \mathbb{C}^{D \times D}$, where $D = 2^N$, we measure a set of Hermitian matrices (observables) $\{\mat{E}_m \in \mathbb{C}^{D \times D}\}_{m=1}^M$, obtaining the expected outcomes $y_m = \operatorname{Tr}(\rho \mat{E}_m)$. The goal of QST is then to reconstruct $\rho$ from the collected data $\{\mat{E}_m, y_m\}_{m=1}^M$. This recovery problem is challenging because (1) $\rho$ must satisfy the physicality constraints, i.e., $\rho \in \mathcal{S}$, and (2) the number of measurement settings required to ensure unique recovery of a full-rank density matrix scales as $\mathcal{O}(D^2)$. However, in the rank-$R$ case, $\mathcal{O}(R D \log^2 D)$ measurement settings suffice to uniquely recover it with high probability \cite{gross2010csqst, liu2011universal, liu2012csqst}.

As mentioned earlier, there are two main formulations in which the low-rank structure can be exploited.
\paragraph*{Rank Minimization}{In this approach, the low-rank constraint is used implicitly as an optimization objective to minimize, with measurement data and physicality as constraints to satisfy. Formally, we can write this as \cite{gross2010csqst}:
	\begin{equation}
		\vspace{-2mm}
		\label{eqn:sdpQST}
		\min_{\hat{\rho} \in \mathbb{C}^{D \times D}} \|\hat{\rho}\|_{\star} \quad \text{s.t.} \quad \|\vec{y} - \ten{M}(\hat{\rho})\|_2 \leq \epsilon, \text{ and } \hat{\rho} \in \mathcal{S}, 
	\end{equation}
	where $\vec{y} \in \mathbb{R}^M$ are the measurement outcomes, and the measurement map $\mathcal{M}$ is $(\mathcal{M}(\rho))_m = \operatorname{Tr}(\mat{E}_m \rho)$. Note that we minimize the nuclear norm (the tightest convex surrogate of the rank function) over a full matrix; hence, it is not a feasible option for mid- to large-scale systems.}

\paragraph*{Error Minimization}{This approach enforces low-rank by parametrizing the density matrix with a fixed, low-rank Cholesky-like (Burer–Monteiro) factorization \cite{burer2003lrsdp}. A differentiable loss is minimized over the low-rank factors, yielding an efficient approximate solution to QST. We can express this as:
	\begin{equation}
		\label{eqn:BMQST}
		\min_{\mat{A} \in \mathbb{C}^{D \times R}} \| \vec{y}  - \ten{M}\left(\hat{\rho}\right) \|_2^2 ~ \text { with } \hat{\rho} = \mat{A}\mat{A}^{\mathrm{H}},\text { s.t. }  \operatorname{Tr}(\hat{\rho})=1. 
	\end{equation}
	Note that the Hermitian and PSD constraints are taken into account implicitly, and the unit-trace constraint is often relaxed to $\|\mathbf{A}\|_{F}^2 \leq 1 \Leftrightarrow \operatorname{Tr}(\hat{\rho}) \leq 1,$ to improve scalability \cite{bhojanapalli2016lrsdp,kyrillidis2018provable}.
}

\paragraph*{Proposed Method} {
	Our approach employs selective QST to directly estimate a structured subset of entries of the density matrix, which can then be leveraged algebraically under low-rank assumptions to reconstruct the full density matrix from this partial information. \par
	As noted earlier, in selective QST, the observables $\mathbf{E}_m$ are chosen so that their expectation values correspond to specific entries of the density matrix \cite{baldwin2016strictly, calderaro2018direct, feng2021direct, morris2019selective, wang2025direct}. Let the Hermitian operators be \(\mathbf{E}^{\mathrm{Re}}_{rc} = \frac{1}{2}(\vec{s}_c \vec{s}_r^\top + \vec{s}_r \vec{s}_c^\top)\) and \(\mathbf{E}^{\mathrm{Im}}_{rc} = \frac{1}{2i}(\vec{s}_c \vec{s}_r^\top - \vec{s}_r \vec{s}_c^\top)\). Thus, we have \(\mathrm{Re}(\oldrho_{rc}) = \operatorname{Tr}(\rho \mathbf{E}^{\mathrm{Re}}_{rc})\) and \(\mathrm{Im}(\oldrho_{rc}) = \operatorname{Tr}(\rho \mathbf{E}^{\mathrm{Im}}_{rc})\), leading to \(\oldrho_{rc} = \overline{\oldrho_{cr}} = \mathrm{Re}(\oldrho_{rc}) + i \, \mathrm{Im}(\oldrho_{rc})\). The diagonal entries \(\oldrho_{rr}\) can be obtained directly from computational-basis measurements, given by \(\oldrho_{rr} = \operatorname{Tr}(\rho \, \vec{s}_r \vec{s}_r^\top)\). More generally, any entry $\oldrho_{rc}$ can be estimated by measuring at most four rank‑1 observables $\{\vec{v}\vec{v}^{\mathrm{H}} \mid \vec{v} \in \mathcal{B}\}$, where $\mathcal{B} = \{\vec{s}_r,\, \vec{s}_c,\, (\vec{s}_r+\vec{s}_c)/\sqrt{2},\, (\vec{s}_r+ i\,\vec{s}_c)/\sqrt{2}\}$. While many other entrywise-probing measurement operators have been studied \cite{baldwin2016strictly, calderaro2018direct, feng2021direct, morris2019selective, wang2025direct}, we use the standard operators  above for theoretical simplicity. Our focus is on determining {\em which} entries to estimate, not on designing measurement operators. 
}

\section{Efficient Low-Rank QST via Structured Measurements}
\label{sec:algqst}
Recent work has shown that structured observation patterns can enable efficient algebraic reconstruction: a generic low‑rank matrix can be uniquely recovered from a certain set of fully observed submatrices, allowing completion via standard NLA operations with deterministic guarantees \cite{bishop2014deterministic, kiray2015mc,  pimentel2016dsc, shakir2024ttfw, stijn2023mlsvdfsj, sofi2025tensor}. Structured patterns for Hermitian PSD completion have likewise been characterized via chordal graphs \cite{Grone1984psdcomp, Smith2008pscomp}, where PSD constraints on the observed principal submatrices are sufficient to ensure a PSD completion of the full state \cite{vandenberghe2015chordal}.

In line with \cite{stijn2023mlsvdfsj, sofi2025tensor}, we identify a set of fully observed submatrices that is informationally complete, meaning that these submatrices uniquely specify the full state. Without loss of generality, we focus on characterizing principal submatrices of the density matrix, as, under mild conditions, they ensure global Hermitian structure and positivity \cite{Grone1984psdcomp, Smith2008pscomp, vandenberghe2015chordal}. Moreover, only the upper (or lower) triangular entries need to be estimated via selective QST. \par
\subsection{Algorithm} \label{sec:alg}
The proposed algorithm consists of two main stages: a measurement step and a reconstruction step. In the measurement step, we select a structured pattern that satisfies the informational-completeness conditions discussed in Section~\ref{subsec:ssi}---\ref{sec:uniq}. This step must also account for the complexity and robustness of implementing the corresponding measurement operators in practical settings. For brevity, we omit circuit-level implementation details and focus on the theoretical aspects. Using the chosen pattern, we obtain structured entries of the underlying density matrix---specifically, principal submatrices---via selective QST or shadow-based tomography \cite{baldwin2016strictly, calderaro2018direct, feng2021direct, morris2019selective, wang2025direct}. In the reconstruction step, we use these (local) submatrices to compute the (global) column space of the density matrix, as discussed in Section~\ref{subsec:ssi}. This column space is then used to recover the full state via a least-squares approach, as outlined in  Section~\ref{subsec:rec}. The pseudocode for the proposed method is provided in Algorithm~\ref{alg:algqst}.

\begin{algorithm}[htb]
	\caption{Algebraic-QST}\label{alg:algqst}
	\Indm
	\KwIn{Selection pattern $\iset{r}_{1}, \ldots, \iset{r}_{L}$ satisfying the conditions of Sec.~\ref{sec:uniq}; rank $R$.\\}
	\KwOut{Estimate of density matrix $\hat{\rho}$}
	\Indp   
	\tcc{{ \small obtain structured density entries}}
	Obtain the principal submatrices $\{\rho_{\mathrm{obs}}^{(l)}\}_{l=1}^{L}$ via direct selective QST or sketching-based QST\;
	\tcc{\small compute orthonormal bases for $\col{\rho}$}
	\For{$l = 1, \ldots, L$}{
		\tcc{\scriptsize compute top-$R$ eigenvectors}
		Compute $\mat{U}^{(l)}, \_, \_  = \texttt{EVD}\left(\rho_{\mathrm{obs}}^{(l)},  R \right)$\; 
		
		Set $\mat{Q}^{(l)} = \left[\mat{S}_{\iset{r}_{l}} \mat{u}^{(l)} \quad \mat{S}_{\iset{r}_{l}^c} \right]$\;
	} 
	Compute $\mat{U}, \_, \_  = \texttt{SVD}\left([\mat{Q}^{(1)}, \ldots, \mat{Q}^{(L)}],  R \right)$\;
	\tcc{{ \small compute reconstruction $\hat{\rho}$}}
	Estimate $\hat{\rho}$ using the submatrices $\{ \rho_{\mathrm{obs}}^{(l)}\}_{l=1}^L$ and the matrix $\mat{U}$, employing either a column-wise reconstruction or an approximate estimation of $\vec{\lambda}$, as described in Section~\ref{subsec:rec}.\;
\end{algorithm}
\unskip 

\subsection{Global Subspace from Local Constraints}\label{subsec:ssi}%
Assume we are given fully observed isorank principal submatrices $\{\rho_{\mathrm{obs}}^{(l)} = \mat{S}_{\iset{r}_{l}}^{\top}\rho \mat{S}_{\iset{r}_{l}}\}_{l=1}^{L}$ of $\rho$ and aim to compute its $R$-dimensional subspace, $\col{\rho}$. The isorank property ensures that these submatrices impose informative constraints on $\col{\rho}$; see \cite{stijn2023mlsvdfsj, sofi2025tensor} for details.  Each $\rho_{\mathrm{obs}}^{(l)}$ admits a compact EVD $\rho^{(l)}_{\mathrm{obs}} = \mat{u}^{(l)} \mat{\Sigma}^{(l)} \mat{u}^{(l) \mathrm{H}}, $ where $\mat{u}^{(l)} \in \C^{|\iset{r}_l| \times R}$ has full column rank $\big(|\iset{r}_l| \geq R \big),$ and $\mat{\Sigma}^{(l)}$ is diagonal with eigenvalues in descending order. Let $Q_l:= \operatorname{span}\left( \mat{Q}^{(l)}\right)$ be the subspace induced by the $l$th submatrix in the ambient (padded) space, with orthonormal basis in $\mat{Q}^{(l)} \!\coloneqq\! \left[\mat{S}_{\iset{r}_{l}} \mat{u}^{(l)} \quad \mat{S}_{\iset{r}_{l}^c}  \right] \in \C^{D \times (R+|\iset{r}_{l}^c|)},$ where $\iset{r}_{l}^{c} \!=\! [D]\setminus \iset{r}_{l}$. Then, we have $\col{\rho} \subseteq \left(\bigcap_{l=1}^L Q_l \right)$. If the submatrices are chosen such that the intersection of the associated subspaces has minimal dimension, $\dim{\left(\bigcap_{l=1}^L Q_l\right)} \!=\! R,$ then
$\col{\rho}\! =\! \bigcap_{l=1}^L Q_l $, yielding unique identifiability. Generic conditions (i.e., conditions that hold with probability 1 when the matrix entries are drawn from a continuous distribution) for this property are given in \cite{stijn2023mlsvdfsj, sofi2025tensor} and will be discussed in Section~\ref{sec:uniq}; we now show how to compute the intersection. \par
To obtain an orthonormal basis $\mat{u}$ for $\col{\rho},$ we estimate a common subspace $U$ closest to $\{Q_l\}_{l=1}^L$:
\begin{equation}
	\label{eqn:subspace}
	U^\star = \argmin_{ \dim(U)=R} \sum_{l=1}^L d_c^2(U, Q_l) \Leftrightarrow \argmax_{\mat{U}^{\mathrm{H}}\mat{U} = \mat{I}_R} \sum_{l=1}^L \left\| \mat{U}^{\mathrm{H}} \left( \mat{q}^{(l)}\right) \right\|_F^2.
\end{equation}
The solution of this problem admits a closed form: $U^\star$ is spanned by the top-$R$ eigenvectors of the sum of projection operators
$\mat{P}_{\mathrm{tot}} \!\coloneqq\! \mat{Q}_{\mathrm{tot}}\mat{Q}_{\mathrm{tot}}^{\mathrm{H}}$,
equivalently by the top-$R$ left singular vectors of the concatenated matrix
$\mat{Q}_{\mathrm{tot}} \!\coloneqq\! [\mat{Q}^{(1)}, \ldots, \mat{Q}^{(L)}]$.
Moreover, $\mat{U}$ can be computed using a matrix--free eigensolver, avoiding the explicit construction of  $\mat{P}_{\mathrm{tot}},$  via its matrix–vector product action:
\begin{equation}
	\label{eqn:matvec}
	v \;\mapsto\;
	\sum_{l=1}^L \Big(
	\mat{S}_{\iset{r}_{l}} \mat{u}^{(l)} \big( \mat{u}^{(l)\mathrm{H}}\,( \mat{S}_{\iset{r}_{l}}^{\top} v) \big)
	\;+\;
	\mat{S}_{\iset{r}_{l}^c} \mat{S}_{\iset{r}_{l}^c}^{\top}\, v
	\Big),
\end{equation}
with per-iteration complexity
$\ten{O}\big(R\sum_{l} |\iset{r}_l|+\sum_{l} |\iset{r}_{l}^c|\big)$. \par

Alternatively, define the effective constraint matrix $\mat{N} \coloneqq L\mat{I}_{\small D}\!-\! \mat{P}_{\mathrm{tot}}$. The effect of the standard basis vectors (corresponding to the missing rows) is annihilated, leaving only the effective subspace constraints; then $\col{\rho}\!=\! \ker(\mat{N})$.

\paragraph*{Error Analysis} Let $\mathbf{E}^{(l)}$ denote the residual error between the best rank-$R$ approximation of the noisy submatrix $\tilde{\rho}^{(l)}_{\mathrm{obs}}$ and the ground-truth  $\rho^{(l)}_{\mathrm{obs}}$, i.e.,  \( \mathbf{E}^{(l)} = \tilde{\rho}^{(l)}_{\mathrm{obs},R} - \rho^{(l)}_{\mathrm{obs}}. \) Let $\tilde{U}_l$ and $U_l$ denote their respective $R$-dimensional column spaces.  Assume (i) the global subspace is uniquely identifiable; (ii) residual errors satisfy \(\|\mat{E}^{(l)}\|_2 \le \epsilon, \; \forall l\); and (iii) a positive spectral gap, \(\lambda_{R} \big(\tilde{\rho}^{(l)}_{\mathrm{obs}}\big)\ge \delta> \epsilon > 0,  \; \forall l\). According to the perturbation bound established by \cite{wedin1972bounds, li2013newbounds}:
\begin{equation}
	\label{eqn:projbound1}
	\|\mat{P}_{\tilde{U}_l} - \mat{P}_{U_l}\|_F
	\le
	\sqrt{2}\,
	\min\!\big\{
	\|\tilde{\rho}^{(l)\dagger}_{\mathrm{obs}, R}\|_2,\,
	\|\rho^{(l)\dagger}_{\mathrm{obs}}\|_2
	\big\}\,
	\|\mat{E}^{(l)}\|_F,
\end{equation}  
and since \(\min\{\|\tilde{\rho}^{(l)\dagger}_{\mathrm{obs}, R}\|_2,\|\rho^{(l)\dagger}_{\mathrm{obs}}\|_2\}\le 1/\delta\) and \(\|\mat{E}^{(l)}\|_F \le \sqrt{|\iset{r}_l|}\,\|\mat{E}^{(l)}\|_2 \le \epsilon\sqrt{|\iset{r}_l|}\), we obtain
\begin{equation}
	\label{eq:projbound2}
	\big\|\mat{P}_{\tilde{U}_l} - \mat{P}_{U_l}\big\|_F
	\;\le\;
	\frac{\epsilon\,\sqrt{2\,|\iset{r}_l|}}{\delta}.
\end{equation}

\begin{figure}[h]
	\centering
	\includegraphics[width=0.13\textwidth]{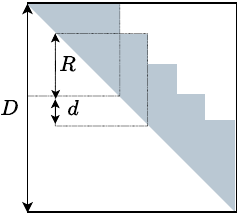}
	\caption{Overlapping block diagonal pattern}
	\label{fig:ovlp}
\end{figure}

Note that \(\|\mat{P}_{\tilde{Q}_l}-\mat{P}_{Q_l}\|_F=\|\mat{P}_{\tilde{U}_l}-\mat{P}_{U_l}\|_F\), since the padded parts are identical, they cancel out in the difference.  Now, suppose the solution is obtained by solving \eqref{eqn:subspace}, i.e., as the (approximate) kernel of $\tilde{\mat{N}} \!=\! L\mat{I}_{\small D}\!-\! \tilde{\mat{P}}_{\mathrm{tot}}$. The error bound between $\tilde{\mat{P}}_{\mathrm{tot}}$ and its noiseless counterpart $\mat{P}_{\mathrm{tot}} $ is given by \vspace{-2mm}
\begin{equation}
	\label{eq:projbound3}
	\|\tilde{\mat P}_{\rm tot}-\mat P_{\rm tot}\|_F^2
	\le
	\Bigl(\sum_{l=1}^{L}\|\mat P_{\tilde Q_l}-\mat P_{Q_l}\|_F\Bigr)^2
	\le
	\frac{2\epsilon^2}{\delta^2}\sum_{l=1}^{L}|\iset r_l|.
\end{equation}
Finally, by applying the results of \cite[Theorem 1.1]{li2013newbounds}  and using $\|\tilde{\mat N}-\mat N\|_F^2
\!=\!\|\tilde{\mat P}_{\rm tot}-\mat P_{\rm tot}\|_F^2$, the chordal distance between the (global) true column space \(U\) and its estimate \(\tilde{U}\) can be bounded as follows\footnote{
We implicitly use the identity
$\|\mathbf{P}_{\tilde U}-\mathbf{P}_{U}\|_F = \|\mathbf{P}_{\tilde U^\perp}-\mathbf{P}_{U^\perp}\|_F$
to obtain the stated subspace bound.
}:
\begin{equation}
	\label{eq:projboundoverall}
	d_c(\tilde{U}, U)
	\;\le\;
	\frac{\epsilon\,\sqrt{2\sum_{l=1}^L |\iset{r}_l|}}
	{\delta\,\sigma_{\min}^+(\tilde{\mat{N}})}.
\end{equation}
The global subspace error is bounded by the local noise-to-signal ratio $\epsilon/\delta$ multiplied by an accumulation factor. If each submatrix contributes $d$ new coordinates (so $|\iset{r}_l| \!=\! R \!+ \!d$), then one possible construction---such as the overlapping block diagonal pattern shown in Fig.~\ref{fig:ovlp}---must satisfy $\cup_{l\!=\!1}^L \iset{r}_l\! =\! [D]$ to ensure coverage of all $D$ coordinates.  See \cite{bishop2014deterministic, pimentel2016dsc, stijn2023mlsvdfsj} for details about these patterns. The orthogonal complement of each isorank submatrix provides $d$  vectors orthogonal to the true $R$-dimensional column space, imposing $2Rd$ real constraints. Since an $R$-dimensional subspace in $\mathbb{C}^D$ has $2R(D\!-\!R)$ intrinsic degrees of freedom, a necessary condition is $L d \!\ge \!(D \!-\! R).$  When $d\!=\!1$, the scheme provides sufficient conditions; for $d\!>\!1$, each submatrix provides more constraints, so fewer are needed. Larger, non-redundant overlaps introduce additional independent constraints in the  matrix $\tilde{\mat{N}}$, tightening the intersection of local subspaces, improving conditioning, and reducing the global error. 

\unskip

\subsection{From Subspace Identification to Full State Reconstruction} \label{subsec:rec}
After computing a basis matrix $\mat{U} \in \mathbb{C}^{D \times R}$, we can recover $\rho$ column-wise. 
Each column $\rho_{:c}$ can be expressed as $\rho_{:c} = \mat{U}\vec{x}$.  If at least $R$ entries of this column are known, we can generically obtain a unique solution $\hat{\vec{x}} \in \mathbb{C}^{R \times 1}$ by solving \( \mat{S}_{\iset{r}_{l}}^{\top} \rho_{:c} = \bigl(\mat{S}_{\iset{r}_{l}}^{\top} \mat{U}\bigr) \vec{x} \) in a least-squares sense. 
Consequently, the estimated column is \( \hat{\rho}_{:c} = \mat{U} \hat{\vec{x}}\).  Alternatively, each entry $\rho_{rc}$ can be written as $\rho_{rc} = (\vec{u}_{r:} \ast \overline{\vec{u}_{c:}})\, \vec{\lambda}$, where $\vec{\lambda}$ is the vector of unknown eigenvalues, which can be estimated in a least-squares sense from the observed entries $(r,c) \in  \cup_{l=1}^L\iset{r}_{l} \times \iset{r}_{l} .$ Consequently, the full density matrix can be reconstructed as $\rho = \mat{U\Lambda U}^{\mathrm{H}},$ where $\mat{\Lambda} = \mathrm{diag}(\vec{\lambda})$.

\subsection{Summary of Necessary and Sufficient Recovery Conditions}
\label{sec:uniq}
In Section~\ref{subsec:ssi}, we showed that the matrix $\mat{U}$, which holds an orthonormal basis for $\col{\rho}$, can be obtained from the isorank submatrices provided the corresponding ambient space subspaces satisfy $\dim \left(\bigcap_{l=1}^L Q_l\right)=R$. This condition is generically satisfied when some submatrices share at least $R$ overlapping rows, i.e., $| \iset{r}_{l_i} \cap \iset{r}_{l_j}| \geq R,$ and every row appears in at least one submatrix,  i.e., $\cup_{l=1}^L \iset{r}_l = [D]$; see \cite{pimentel2016dsc, stijn2023mlsvdfsj, sofi2025tensor}. Consequently, to recover the full density matrix uniquely, every column (resp. row) must contain at least $R$ known entries, as discussed in Section~\ref{subsec:rec}.  These conditions are not only sufficient for computation but also necessary for uniqueness. The structured pattern shown in Fig.~\ref{fig:ovlp} satisfies these conditions generically.

\paragraph*{Measurement Complexity}{A rank-$R$ $D \times D$ density matrix has $2RD \!-\! R^{2}\!-\!1$ real parameters, implying that at least this many measurement are necessary for QST. Under random Pauli measurement settings, the sample complexity scales as $\mathcal{O}(R D \log^2 D)$ instead of $\mathcal{O}(R D)$, due to the coupon-collector–like effect \cite{gross2010csqst, candes2010power}. For the structured pattern shown in Fig.~\ref{fig:ovlp}, selective QST requires $M = (R\!+\!d)^2 +  (L\!-\!1)\big((R+d)^2-R^2\big)$ distinct measurement settings. When $d\!=\!1$, we have $L = D - R,$ and the overall complexity scales as $\mathcal{O}(RD)$, for $R \le D.$ Specifically,  for the rank-$1$ case, $M$ reduces to $3D - 2,$ consistent with the result of  \cite{finkelstein2004pure}.
}

\subsection{\label{sec:shadowSQST} Classical Shadows for Structured Density Matrix Entries}
In standard selective QST, each entry requires a separate measurement, precluding reuse of prior data. In contrast, shadow tomography constructs classical shadows from \(\mathcal{O}\!\left(\frac{\log(M/\delta)}{\varepsilon^{2}}\right)\) random Clifford or (biased) mutually unbiased unitaries and reuses them to estimate \(M\) entries within additive error \(\varepsilon\) and confidence \(1\!-\!\delta\) \cite{aaronson2018shadow, huang2020predicting, morris2019selective, wang2025direct}. Hence, using such an approach to obtain structured entries could yield significant scalability gains.
\section{Experiments}
\label{sec:exp}
Accuracy is measured by fidelity $\left( \operatorname{Tr}\left(\sqrt{\sqrt{\rho} \hat{\rho} \sqrt{\rho}} \right) \right)^2$ and trace distance $\tfrac{1}{2}\|\rho-\hat{\rho}\|_{1}$.
Experiments are conducted on an HP EliteBook 845 G8 with an AMD Ryzen 7 PRO 5850U processor and 32GB RAM.

\subsection{Results}
A random rank-\(2\) density matrix of an \(N\)-qubit system is generated from the Ginibre ensemble for \(N \!=\! 4, 5,\) and \(6\). Standard selective QST is used to obtain a structured block-diagonal measurement pattern, as shown in Fig.~\ref{fig:ovlp}. Here, \(d\) is varied from \(1\) to \(5\), making overlapping blocks of size \(R \!+\! d\) with overlap \(R\), thereby ensuring that the conditions of Section~\ref{sec:uniq} are generically satisfied. Gaussian noise is added to the measurements with a signal-to-noise ratio of \(30\,\mathrm{dB}\).

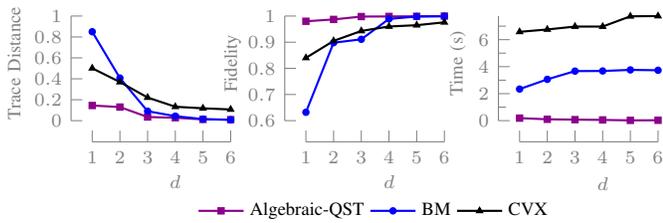
\begin{figure}[h]
	\centering
	\input{figures/3b3resfig}
	\caption{The proposed method yields higher accuracy than both references while also enabling significantly faster reconstruction. Moreover, as $d$ grows, more measurements become available per parameter (with $D$ and $R$ fixed), leading to improved accuracy.  For fixed \(R\), \(d\) (especially when $d$ is small), and a constant noise level, increasing \(N\) slightly weakens the subspace perturbation bound due to its dependence on the ratio of \( L \propto D\!-\!R\) to  $\sigma_{\min}^+(\tilde{\mat{N}})$; see Eqn.~\eqref{eq:projboundoverall}. We observe in our experiments that this ratio increases; consistently, we see a slight increase in subspace error and, consequently, in reconstruction error. }
	\label{fig:resfig}
	\vspace{-1em}
\end{figure}

The proposed algebraic algorithm is compared with SOTA methods, including a convex optimization (CVX) approach~\cite{gross2010csqst, liu2011universal, liu2012csqst}, which solves the SDP in~\eqref{eqn:sdpQST}, and a low-rank Burer--Monteiro (BM) factorization method~\cite{burer2003lrsdp, bhojanapalli2016lrsdp, kyrillidis2018provable}, which solves a relaxed SDP in~\eqref{eqn:BMQST} via second-order optimization. In the latter two approaches, the measurement operators are random $N$-qubit Pauli operators $\mat{E}_m := \otimes_{j=1}^{N} \bm{\sigma}_{m_j}$ with $\bm{\sigma}_{m_j} \in \{\mat{I}_2,\bm{\sigma}_x,\bm{\sigma}_y,\bm{\sigma}_z\}$, and the same noise level is applied. The number of measurements equals twice the unique off-diagonal entries plus $D$ diagonals, ensuring the same number of measurements across all methods. Median fidelity and trace distance over 15 trials are reported in Fig.~\ref{fig:resfig}.

\label{sec:res}
\unskip

\section{Conclusion}
\label{sec:con}
An algebraic QST approach is introduced for reconstructing generic low-rank quantum states, combining structured matrix completion and selective QST, with deterministic recovery guarantees. Numerical experiments show that the method yields accurate, valid results and is computationally fast. The estimation of structured entries via shadow tomography is briefly noted; future work will provide more detailed numerical experiments, show shadow-based estimation, and discuss practical (circuit-level) implementations.
\printbibliography
\end{document}

%% file: figures/3b3resfig.tex
\begin{tikzpicture}[scale=0.45]
	
	\begin{groupplot}[
		group style={group size=3 by 3, horizontal sep=1.75cm, vertical sep=1.2cm},
		width=0.3\textwidth,
		height=0.25\textwidth,
		grid=both,
		legend to name=sharedleg,
		legend style={draw=none, legend image post style={line width=1pt, mark size=1pt,}},
		legend columns=3,
		every axis x label/.append style={
			at={(axis description cs:0.5,-0.35)},
			anchor=north, color=black!10!gray,
		},
		every axis y label/.append style={
			at={(axis description cs:-0.4,0.5)},
			anchor=south,
		 color=black!10!gray,
		 },
		x axis line style={yshift=-8pt,-,color=gray,},
		y axis line style={xshift=-8pt,-,color=gray,},
		xticklabel style={yshift=-8pt,color=gray, 	font=\Large},
		yticklabel style={xshift=-8pt,color=gray, font=\Large},
		xtick style={yshift=-8pt, line width=0.3pt, 	font=\Large},
		ytick style={xshift=-8pt, line width=0.3pt, 	font=\Large},
		xtick align=inside,
		ytick align=inside,
		]

		
		\nextgroupplot[
		ylabel={\Large Fidelity},
		title={\color{black!10!gray} \Large Algebraic-QST},
		axis lines=left,
		xtick={1,2,3,4,5},
		ymin=0.90, ymax=1.,
		]
		
		\addplot[ultra  thick, olive, mark=square*, mark size=2pt]
		coordinates {(1,0.9629) (2,0.9799) (3,0.9995) (4,0.9997)(5, 1)};
		\addplot[ultra  thick, teal, mark=*, mark size=2pt]
		coordinates {(1,0.9626) (2,0.9887) (3,0.9998) (4,0.9996)(5,0.9997)};
		\addplot[ultra  thick, black, mark=triangle*, mark size=2pt]
		coordinates {(1,0.9592) (2,0.9784) (3,0.9952) (4,1.0)(5, 0.99862)};
		
		\nextgroupplot[
		title={\color{black!10!gray}  \Large CVX},
		axis lines=left,
		xtick={1,2,3,4, 5},
		ymin=0.5, ymax=1.0,
		]
		
\addplot[ultra  thick, olive, mark=square*, mark size=2pt]
coordinates {(1,0.8944) (2,0.9398) (3,0.9706) (4,0.982)(5, 0.9915 )};
\addplot[ultra  thick, teal, mark=*, mark size=2pt]
coordinates {(1,0.8137) (2,0.8632) (3,0.9275) (4,0.9614)(5,0.9661)};
\addplot[ultra  thick, black, mark=triangle*, mark size=2pt]
coordinates {(1,0.7282) (2,0.8292) (3,0.9078) (4,0.9503)(5, 0.9631)};
		
		\nextgroupplot[
		title={\color{black!10!gray}  \Large BM},
		axis lines=left,
		xtick={1,2,3,4, 5},
		ymin=0.4, ymax=1.,
		]
		
		\addplot[ultra  thick, olive, mark=square*, mark size=2pt]
		coordinates {(1,0.7239) (2,0.916) (3,0.9987) (4,0.9991)(5, 0.9995 )};
		\addplot[ultra  thick, teal, mark=*, mark size=2pt]
		coordinates {(1,0.687) (2,0.8134) (3,0.9721) (4,0.9989)(5,0.9992)};
		\addplot[ultra  thick, black, mark=triangle*, mark size=2pt]
		coordinates {(1,0.5617) (2,0.623)(3,0.7528) (4,0.9255)(5, 0.9571)};
		
		
		\nextgroupplot[
		ylabel={\Large Trace Distance},
		axis lines=left,
		xtick={1,2,3,4, 5},
		ymin = 1e-2,
		ymax=1,
		ymode=log,
		ytick={1e-2,1e-1,1e0},
		minor tick num=0,
		]
		
		\addplot[ultra  thick, olive, mark=square*, mark size=2pt]
		coordinates {(1,0.1236) (2,0.1033) (3,0.0317) (4,0.0168)(5, 0.0193)};
		\addplot[ultra  thick, teal, mark=*, mark size=2pt]
		coordinates {(1,0.2946) (2,0.1224) (3,0.0791) (4,0.05)(5,0.0317)};
		\addplot[ultra  thick, black, mark=triangle*, mark size=2pt]
		coordinates {(1,0.3723) (2,0.2306) (3,0.187) (4,0.0758)(5,0.0685)};
		
		\nextgroupplot[
		axis lines=left,
		xtick={1,2,3,4,5},
		ymin = 1e-2,
		ymax=1,
		ymode=log,
		ytick={1e-2,1e-1,1e0},
		]
		
		\addplot[ultra  thick, olive, mark=square*, mark size=2pt]
		coordinates {(1,0.2966) (2,0.1826) (3,0.1171) (4,0.0941)(5, 0.0907)};
		\addplot[ultra  thick, teal, mark=*, mark size=2pt]
		coordinates {(1,0.415) (2,0.3278) (3,0.2013) (4,0.137)(5, 0.1252)};
		\addplot[ultra  thick, black, mark=triangle*, mark size=2pt]
		coordinates {(1,0.5438) (2,0.3809) (3,0.2445) (4,0.1618)(5, 0.1337)};
		
		\nextgroupplot[
		axis lines=left,
		xtick={1,2,3,4,5},
		ymin = 1e-2,
		ymax=1,
		ymode=log,
		ytick={1e-2,1e-1,1e0},
		]
		
		\addplot[ultra  thick, olive, mark=square*, mark size=2pt]
		coordinates {(1,0.5818) (2,0.2225) (3,0.0473) (4,0.0384)(5, 0.0299)};
		\addplot[ultra  thick, teal, mark=*, mark size=2pt]
		coordinates {(1,0.5698) (2,0.5056) (3,0.1158) (4,0.0456)(5, 0.0385)};
		\addplot[ultra  thick, black, mark=triangle*, mark size=2pt]
		coordinates {(1,0.7806) (2,0.746) (3,0.4826) (4,0.2857)(5,0.16)};

		
		\nextgroupplot[
		ylabel={\Large Time (s)},
		axis lines=left,
		ymode=log,
		ymin=1e-3,
		ymax=1e1,
		xtick={1,2,3,4, 5},
		xlabel={\Large $d$},
		ytick={1e-3, 1e-2, 1e-1, 1e0, 1e1},
		]
		
		\addplot[ultra  thick, olive, mark=square*]
		coordinates {(1,0.0069) (2,0.0039) (3,0.0015) (4,0.0013)(5, 0.001)};
		\addplot[ultra  thick, teal, mark=*]
		coordinates {(1,0.0445) (2,0.0181) (3,0.013) (4,0.0126)(5, 0.0123)};
		\addplot[ultra  thick, black, mark=triangle*]
		coordinates {(1,1.75023) (2,0.5958) (3,0.2274) (4,0.0617) (5, 0.0518)};
		
		\nextgroupplot[
		axis lines=left,
		ymode=log,
		ymin=1e-1,
		ymax=1e2,
		xtick={1,2,3,4, 5},
		xlabel={\Large $d$},
		ytick={1e-1,1e0,1e1,1e2},
		]
		
		\addplot[ultra  thick, olive, mark=square*, mark size=2pt]
		coordinates {(1,1.3108) (2,1.3336) (3,1.402) (4,1.4493)(5,1.5155)};
		\addplot[ultra thick, teal, mark=*, mark size=2pt]
		coordinates {(1,6.323) (2,6.735) (3,6.99) (4,6.604)(5,6.8077)};
		\addplot[ultra thick, black, mark=triangle*, mark size=2pt]
		coordinates {(1,29.564) (2,32.5261) (3,33.6197) (4,34.7065)(5,36.8742)};
		
		\nextgroupplot[
		axis lines=left,
		ymode=log,
		ymin=1e-1,
		ymax=1e2,
		xtick={1,2,3,4, 5},
    	ytick={1e-1,1e0,1e1,1e2},
		xlabel={\Large $d$},
		]
		
		\addplot[ultra thick, olive, mark=square*, mark size=2pt]
		coordinates {(1,1.1819) (2,0.9616) (3,0.78) (4,0.7242)(5,0.6779)};
		\addplot[ultra thick, teal, mark=*, mark size=2pt]
		coordinates {(1,5.7368) (2,6.6153) (3,4.6011) (4,4.0414)(5,3.4855)};
		\addplot[ultra thick, black, mark=triangle*, mark size=2pt]
		coordinates {(1,9.1041) (2,13.5061) (3, 19.7335) (4,22.9682)(5, 26.507)};

		\addlegendentry{\color{black!10!gray} \scriptsize $N=4$}
		\addlegendentry{\color{black!10!gray} \scriptsize  $N=5$}
		\addlegendentry{\color{black!10!gray} \scriptsize  $N=6$}
		
	\end{groupplot}
	
	\node at ($(group c2r3.south) + (0,-2.1cm)$)
	{\pgfplotslegendfromname{sharedleg}};
	
\end{tikzpicture}

%% file: ref.bib
@article{eisert2010colloquium,
  title={Colloquium: Area laws for the entanglement entropy},
  author={Eisert, Jens and Cramer, Marcus and Plenio, Martin B},
  journal={Rev. Mod. Phys.},
  volume={82},
  number={1},
  pages={277--306},
  year={2010},
  publisher={APS}
}

@article{pimentel2016dsc,
  author={Pimentel-Alarcón, Daniel L. and Boston, Nigel and Nowak, Robert D.},
  title={A Characterization of Deterministic Sampling Patterns for Low-Rank Matrix Completion},
  year={2016},
  journal={IEEE J. Sel. Top. Signal Process.}, 
  volume={10},
  pages={623-636},
}

@article{oseledets2010tensortrain,
  ids =		 {oseledets2011tensortraindecomposition},
  author =	 {I. Oseledets},
  title =	 {Tensor-train decomposition},
  year =	 2011,
  journal =	 {SIAM J. Sci. Comput.},
  volume =	 33,
  pages =	 {2295-2317}
}

@article{verstraete2007MPS,
author = {Perez-Garcia, D. and Verstraete, F. and Wolf, M. M. and Cirac, J. I.},
title = {Matrix product state representations},
year = {2007},
journal = {Quantum Inf. Comput.},
volume = {7},
pages = {401–430},
}

@book{nielsen2010quantum,
  title={Quantum computation and quantum information},
  author={Nielsen, Michael A and Chuang, Isaac L},
  year={2010},
  publisher={Cambridge university press}
}

@article{gross2010csqst,
  title = {Quantum State Tomography via Compressed Sensing},
  author = {Gross, David and Liu, Yi-Kai and Flammia, Steven T. and Becker, Stephen and Eisert, Jens},
  journal = {Phys. Rev. Lett.},
  volume = {105},
  issue = {15},
  pages = {150401},
  numpages = {4},
  year = {2010},
  month = {10},
  publisher = {American Physical Society},
}

@article{liu2011universal,
  title={Universal low-rank matrix recovery from Pauli measurements},
  author={Liu, Yi-Kai},
  journal={Adv. Neural Inf. Process. Syst.},
  volume={24},
  year={2011}
}

@article{liu2012csqst,
  title = {Experimental Quantum State Tomography via Compressed Sampling},
  author = {Liu, Wei-Tao and Zhang, Ting and Liu, Ji-Ying and Chen, Ping-Xing and Yuan, Jian-Min},
  journal = {Phys. Rev. Lett.},
  volume = {108},
  issue = {17},
  pages = {170403},
  numpages = {5},
  year = {2012},
  publisher = {American Physical Society},
}

@article{burer2003lrsdp,
  title={A nonlinear programming algorithm for solving semidefinite programs via low-rank factorization},
  author={Samuel Burer and Renato D. C. Monteiro},
  journal={Math. Program.},
  year={2003},
  volume={95},
  pages={329-357},
}

@article{kyrillidis2018provable,
  title={Provable compressed sensing quantum state tomography via non-convex methods},
  author={Kyrillidis, Anastasios and Kalev, Amir and Park, Dohyung and Bhojanapalli, Srinadh and Caramanis, Constantine and Sanghavi, Sujay},
  journal={npj Quantum Inf.},
  volume={4},
  number={1},
  pages={36},
  year={2018},
}

@inproceedings{bhojanapalli2016lrsdp,
  title={Dropping convexity for faster semi-definite optimization},
  author={Bhojanapalli, Srinadh and Kyrillidis, Anastasios and Sanghavi, Sujay},
  booktitle={Conference on Learning Theory},
  pages={530--582},
  year={2016},
  organization={PMLR}
}

@article{cramer2010efficient,
  title={Efficient quantum state tomography},
  author={Cramer, Marcus and Plenio, Martin B and Flammia, Steven T and Somma, Rolando and Gross, David and Bartlett, Stephen D and Landon-Cardinal, Olivier and Poulin, David and Liu, Yi-Kai},
  journal={Nat. Commun.},
  volume={1},
  number={1},
  pages={149},
  year={2010},
}

@article{lanyon2017efficient,
  title={Efficient tomography of a quantum many-body system},
  author={Lanyon, Ben P and Maier, Christine and Holz{\"a}pfel, Milan and Baumgratz, Tillmann and Hempel, Cornelius and Jurcevic, Petar and Dhand, Ish and Buyskikh, AS and Daley, Andrew J and Cramer, Marcus and others},
  journal={Nat. Phys.},
  volume={13},
  number={12},
  pages={1158--1162},
  year={2017},
}

@article{stijn2023mlsvdfsj,
author = {S\o{}rensen, Mikael and Hendrikx, Stijn and De Lathauwer, Lieven},
title = {Multilinear Singular Value Decomposition–Based Completion with Fibers Observed in a Single Mode},
journal = {SIAM J. Matrix Anal. Appl.},
volume = {46},
number = {2},
pages = {1061-1090},
year = {2025},
}

@inproceedings{shakir2024ttfw,
	title={Tensor Train Completion of Multi-Way Data Observed Along One Mode},
	author={Sofi, Shakir Showkat and Hendrikx, Stijn and De Lathauwer, Lieven},
	booktitle={Proceedings of the 32nd EUSIPCO},
	year={2024},
        pages={1067-1071},
	organizers = {IEEE}
}

@Article{sofi2025tensor,
AUTHOR = {Sofi, Shakir Showkat and De Lathauwer, Lieven},
TITLE = {Tensor Train Completion from Fiberwise Observations Along a Single Mode},
JOURNAL = {Mathematics},
VOLUME = {14},
YEAR = {2026},
NUMBER = {5},
ARTICLE-NUMBER = {922},
URL = {https://www.mdpi.com/2227-7390/14/5/922},
ISSN = {2227-7390},
DOI = {10.3390/math14050922}
}

@article{li2013newbounds,
author = {Li, Bingxiang and Li, Wen and Cui, Lubin},
title = {New bounds for perturbation of the orthogonal projection},
year = {2013},
issue_date = {March 2013},
publisher = {Springer-Verlag},
address = {Berlin, Heidelberg},
volume = {50},
number = {1},
issn = {0008-0624},
journal = {Calcolo},
month = mar,
pages = {69–78},
numpages = {10},
}

@INPROCEEDINGS{sofi2025bttqst,
  author={Sofi, Shakir Showkat and Vermeylen, Charlotte and De Lathauwer, Lieven},
  booktitle={2025 33rd European Signal Processing Conference (EUSIPCO)}, 
  title={Tensor Train Quantum State Tomography Using Compressed Sensing}, 
  year={2025},
  volume={},
  number={},
  pages={1332-1336},
  doi={10.23919/EUSIPCO63237.2025.11226412}}

@article{toth2010permutationally,
  title={Permutationally invariant quantum tomography},
  author={T{\'o}th, G{\'e}za and Wieczorek, Witlef and Gross, David and Krischek, Roland and Schwemmer, Christian and Weinfurter, Harald},
  journal={Phys. Rev. Lett.},
  volume={105},
  number={25},
  pages={250403},
  year={2010},
  publisher={APS}
}

@article{moroder2012permutationally,
  title={Permutationally invariant state reconstruction},
  author={Moroder, Tobias and Hyllus, Philipp and T{\'o}th, G{\'e}za and Schwemmer, Christian and Niggebaum, Alexander and Gaile, Stefanie and G{\"u}hne, Otfried and Weinfurter, Harald},
  journal={New J. Phys.},
  volume={14},
  number={10},
  pages={105001},
  year={2012},
  publisher={IOP Publishing}
}

@inproceedings{aaronson2018shadow,
  title={Shadow tomography of quantum states},
  author={Aaronson, Scott},
  booktitle={Proceedings of the 50th annual ACM SIGACT symposium on theory of computing},
  pages={325--338},
  year={2018}
}

@article{huang2020predicting,
  title={Predicting many properties of a quantum system from very few measurements},
  author={Huang, Hsin-Yuan and Kueng, Richard and Preskill, John},
  journal={Nat. Phys.},
  volume={16},
  number={10},
  pages={1050--1057},
  year={2020},
  publisher={Nature Publishing Group UK London}
}

@article{baldwin2016strictly,
  title={Strictly-complete measurements for bounded-rank quantum-state tomography},
  author={Baldwin, Charles H and Deutsch, Ivan H and Kalev, Amir},
  journal={Phys. Rev. A},
  volume={93},
  number={5},
  pages={052105},
  year={2016},
  publisher={APS}
}

@article{calderaro2018direct,
  title={Direct reconstruction of the quantum density matrix by strong measurements},
  author={Calderaro, Luca and Foletto, Giulio and Dequal, Daniele and Villoresi, Paolo and Vallone, Giuseppe},
  journal={Phys. Rev. Lett.},
  volume={121},
  number={23},
  pages={230501},
  year={2018},
  publisher={APS}
}

@article{feng2021direct,
  title={Direct measurement of density-matrix elements using a phase-shifting technique},
  author={Feng, Tianfeng and Ren, Changliang and Zhou, Xiaoqi},
  journal={Phys. Rev. A},
  volume={104},
  number={4},
  pages={042403},
  year={2021},
  publisher={American Physical Society}
}

@article{morris2019selective,
  title={Selective quantum state tomography},
  author={Morris, Joshua and Daki{\'c}, Borivoje},
  journal={arXiv preprint arXiv:1909.05880},
  year={2019}
}

@article{wang2025direct,
  title={Direct reconstruction of the quantum density matrix elements with classical shadow tomography},
  author={Wang, Yu},
  journal={arXiv preprint arXiv:2505.15243},
  year={2025}
}

@article{Grone1984psdcomp,
title = {Positive definite completions of partial Hermitian matrices},
journal = {Linear Algebra Appl.},
volume = {58},
pages = {109-124},
year = {1984},
issn = {0024-3795},
author = {Robert Grone and Charles R. Johnson and Eduardo M. Sá and Henry Wolkowicz},
}

@article{Smith2008pscomp,
title = {The positive definite completion problem revisited},
journal = {Linear Algebra Appl.},
volume = {429},
number = {7},
pages = {1442-1452},
year = {2008},
issn = {0024-3795},
author = {Ronald L. Smith},
}

@inproceedings{bishop2014deterministic,
 author = {Bishop, William E. and Yu, Byron M.},
 booktitle = {Advances in Neural Information Processing Systems},
 editor = {Z. Ghahramani and M. Welling and C. Cortes and N. Lawrence and K.Q. Weinberger},
 pages = {},
 publisher = {Curran Associates, Inc.},
 title = {Deterministic Symmetric Positive Semidefinite Matrix Completion},
 volume = {27},
 year = {2014}
}

@article{vandenberghe2015chordal,
  title={Chordal graphs and semidefinite optimization},
  author={Vandenberghe, Lieven and Andersen, Martin S},
  journal={Found. Trends Optim.},
  volume={1},
  number={4},
  pages={241--433},
  year={2015},
  publisher={Emerald Publishing Limited}
}

@article{kiray2015mc,
  author  = {Franz J.Kir{{\'a}}ly and Louis Theran and Ryota Tomioka},
  title   = {The Algebraic Combinatorial Approach for Low-Rank Matrix Completion},
  journal = {J. Mach. Learn. Res.},
  year    = {2015},
  volume  = {16},
  pages   = {1391-1436},
}

@article{wedin1972bounds,
author = {Wedin, P},
title = {Perturbation bounds in connection with singular value decomposition},
year = {1972},
issue_date = {Mar 1972},
publisher = {BIT Numer. Math.},
address = {USA},
volume = {12},
number = {1},
issn = {0006-3835},
journal = {BIT},
month = mar,
pages = {99–111},
numpages = {13},
}

@article{candes2010power,
  title={The power of convex relaxation: Near-optimal matrix completion},
  author={Cand\`{e}s, Emmanuel J and Tao, Terence},
  journal={IEEE Trans. Inf. Theory},
  volume={56},
  number={5},
  pages={2053--2080},
  year={2010},
  publisher={IEEE}
}

@article{finkelstein2004pure,
  title={Pure-state informationally complete and “really” complete measurements},
  author={Finkelstein, Jerome},
  journal={Phys. Rev. A},
  volume={70},
  number={5},
  pages={052107},
  year={2004},
  publisher={APS}
}
